\documentclass[aps,pra,superscriptaddress,twocolumn,groupedaddress]{revtex4-1}
\usepackage{mathrsfs}
\usepackage{amsmath}
\usepackage{amssymb}
\usepackage{amstext}
\usepackage{graphicx}
\usepackage{color}
\usepackage{epstopdf}

\begin{document}


\title{Numerical evaluation of the fidelity error threshold for the
  surface code}

\author{Pejman Jouzdani}

\affiliation{Department of Physics, University of Central Florida,
  P.O. Box 162385, Orlando, Florida 32816, USA}

\author{Eduardo R. Mucciolo}

\affiliation{Department of Physics, University of Central Florida,
  P.O. Box 162385, Orlando, Florida 32816, USA}




\date{\today}

\begin{abstract}
We study how the resilience of the surface code is affected by the
coupling to a non-Markovian environment at zero temperature. The
qubits in the surface code experience an effective dynamics due to the
coupling to the environment that induces correlations among them. The
range of the effective induced qubit-qubit interaction depends on
parameters related to the environment and the duration of the quantum
error correction cycle. We show numerically that different interaction
ranges set different intrinsic bounds on the fidelity of the
code. These bounds are unrelated to the error thresholds based on
stochastic error models. We introduce a definition of stabilizers
based on logical operators that allows us to efficiently implement a
Metropolis algorithm to determine upper bounds to the fidelity error
threshold.
\end{abstract}


\maketitle


\section{Introduction}

Topological quantum codes provide an alternative route to
fault-tolerance quantum computation. In topological quantum codes the
information is encoded on the topological characteristics of the
physical system, resulting in protection against local perturbations
\cite{Kitaev03,Dennis02,Freedman03,Nayak08}. The surface (or planar)
code \cite{Sergey98,Freedman98} is an important example of this class
of quantum codes that requires an active approach to error correction
on lattice of regular qubits. In contrast to the toric code
\cite{Kitaev03}, which has an intrinsic Hamiltonian that governs the
evolution of the system, the surface code has no intrinsic Hamiltonian
and the system's evolution is due to its coupling to an environment
and the syndrome extraction and recovery operations at the end of a
cycle. It has attracted increasing attention in recent years due to
its more practical nature than other topological forms of
encodings. Architectures based on superconducting qubits
\cite{Divicenzo09} and Majorana fermions \cite{Terhal12} have been
proposed theoretically. At the experimental level, significant
increase in coherence time and fidelity of logical gates in
superconducting qubits has been recently reported \cite{Hanhee11,
  Martinis05, Barends14}, suggesting that these systems may provide a
suitable experimental setting for implementing surface codes. Several
studies have been done to determine the error threshold of
two-dimensional topological codes \cite{Raussendorf07, Fowler09,
  Wootton12, Hutter14, Terhal13, Bombin12}. However, in these studies
the role played by correlated errors was not fully
investigated. However, it is crucial to study the impact of correlated
errors on any scalable quantum code before attempting to quantify
error thresholds based on quantum error correction (QEC) protocols
\cite{Preskill12}.

When in contact with environmental degrees of freedom, the physical
qubits in the surface code will experience an effective dynamics. This
effective dynamics may comprise qubit-qubit interactions, which in
turn can result in a correlated time evolution. Since a large-scale
quantum code has a large Hilbert space, a correlated dynamics may
cause a sharp change in the quantum phase of the code system, even in
the presence of QEC operations. This change of phase cannot be studied
in the context of stochastic noise models, which typically only
include bit flip, phase flip, and depolarizing channels. 

The effective dynamics induced by the environment on the code system
is in general very nontrivial to derive from first
principles. However, for a particular case, the bosonic bath, we were
able to obtain an exact effective action after a single QEC cycle
\cite{novais2013,jouzdani2013}. This action comprises a qubit-qubit
interaction term with a distance-dependent exchange coupling. The
range and strength of qubit-qubit interaction were found to be
functions of environmental parameters, the distance between the
qubits, and the duration of the QEC cycle. The effective dynamics
derived for a bosonic bath could be used as a phenomenological model
for other types of environments as it has a rather general functional
form.

In this paper we numerically evaluate the effect of correlated errors
induced by a two-qubit effective action and study the impact of
different ranges of correlations. We use a Monte Carlo method for
evaluating the fidelity of the surface code at the end of a complete
QEC cycle. We introduce an alternative approach to define the surface
code stabilizers that helps us to implement an efficient Metropolis
algorithm. This method can be extended to other topological systems
such as the toric code. For the surface code, we confirm the results
presented in Ref. \cite{jouzdani14}, namely, the existence of a sharp
transition in the fidelity as a function of the coupling between
qubits and the environment for large enough codes. The critical value
of this coupling provides a threshold for the ability of the surface
code to protect quantum information. We also find that an increase in
the correlation range does not wash away this critical point but
moves it to lower coupling constant values, making it more difficult
in practice to achieve protection. 

The paper is organized as follows. In Sec. \ref{surf} we introduced
the basic elements of the surface code. In Sec. \ref{sec:environment}
we describe a model interaction for the environment and the physical
qubits that induces an effective evolution for the code system. We
then use this evolution in Sec. \ref{QEC} to obtain an expression for
the surface code fidelity in terms of expectation values of a spin
statistical model. The numerical Monte Carlo method used to compute
these expectation values and the results are described in
Secs. \ref{sec:method} and \ref{sec:results}. Finally, a summary is
provided in Sec. \ref{sec:summary}.

\section{Surface Code}
\label{surf}

\begin{figure}[t]
\includegraphics[width=0.8\columnwidth]{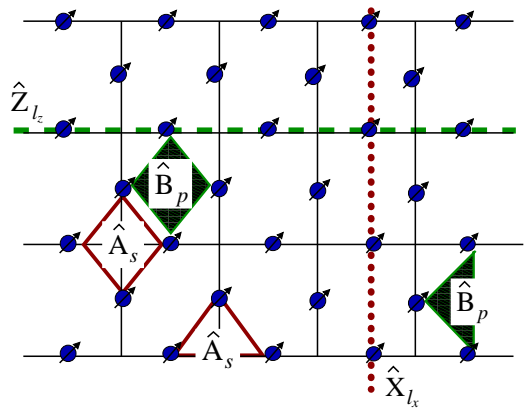}
\caption{(Color online) The geometry of the surface code
  system. Physical qubits are shown with arrows. A plaquette (star)
  operator $\hat{B}_{p}$ ($\hat{A}_{s}$) is shown with a shaded
  (unshaded) enclosed area connecting the corresponding qubits. The
  dashed green (dotted red) line $l_z$ ($l_x$) represents the logical
  operator $\hat{Z}_{l_z}$($\hat{X}_{l_x}$) and runs over
  corresponding qubits. }
\label{fig:SC1}
\end{figure}

The surface code \cite{Sergey98, Freedman98} is a collection of $N$
qubits located on the links of a two-dimensional lattice, as shown in
Fig. \ref{fig:SC1}. There are two types of stabilizers, $\hat{A}_s$
and $\hat{B}_p$, which are defined as
\begin{eqnarray}
  \hat{A}_s= \prod_{i\in s} \sigma_i^x
  \label{eq:StabilizersA}
\end{eqnarray}
and
\begin{eqnarray}
  \hat{B}_p= \prod_{i\in p} \sigma_i^z.
  \label{eq:StabilizersB}
\end{eqnarray}
They have eigenvalues $A_s$ and $B_p$, respectively, that take the
values $\pm1$. The subscript $s$($p$) refers to a vertex (plaquette)
on the lattice, and $\sigma_i^\alpha$ is the $\alpha$ component of the
Pauli matrices that acts on the $i$-th qubit. The two stabilizers
commute, $[\hat{A}_s, \hat{B}_p]=0$, and thus are simultaneously
observable.  In addition, there are two logical operators defined as
\begin{eqnarray}
\hat{X}_{l_x}= \prod_{i\in l_x} \sigma_i^x,
\label{eq:logicalX}
\end{eqnarray}
and
\begin{eqnarray}
  \hat{Z}_{l_z}= \prod_{i\in l_z} \sigma_i^z,
  \label{eq:logicalZ}
\end{eqnarray}
where the path $l_x$ ($l_z$) runs from one boundary to the opposite
boundary, as shown in Fig. \ref{fig:SC1}. The two logical
operators follow the same commutation relations as the Pauli matrices
$\sigma^x$ and $\sigma^z$ and both commute with the stabilizers in
Eqs. (\ref{eq:StabilizersA}) and (\ref{eq:StabilizersB}).

The \emph{code space} is defined as the particular subspace of the
total Hilbert space of the system for which the outcome of any
stabilizer is $+1$. The maximum set of observables can be either
$ \text{\bf\{}   \{\hat{A}\},\{\hat{B} \} ,\hat{Z}\text{\bf\}}$ or
$\text{\bf\{}\{\hat{A}\},\{\hat{B}\},\hat{X}\text{\bf\}}$. Considering the set
$\text{\bf\{} \{\hat{A}\},\{\hat{B}\},\hat{Z}\text{\bf\}}$, $\hat{X}$ anticommutes with the
logical operator in the set, $\hat{Z}$. Therefore, there are only two
distinct basis states $|SC\rangle$ and $\hat{X}|SC\rangle$ that
satisfy the condition of the code space.  $\hat{X}$ can be chosen
along different paths $l_x$. However $\hat{X}|SC\rangle$ is unique and
independent of the chosen logical operator. These two orthogonal
states are the two states of the logical qubit of the surface code.
The code is topologically protected, e.g., to flip the logical qubit
state $|SC\rangle$ to $\hat{X}|SC\rangle$ a logical operator
comprising at least $L$ physical qubit operations must be applied,
where $L$ is the linear size of the system. We will refer to
$|SC\rangle$ as the \emph{code state}.

Any deviation from the code space due to local errors such as qubit
flips or phase flips results in excitations known as anyons. The
anyons correspond to stabilizers that yield an outcome $-1$ after
measurement.  Anyons corresponding to $\hat{A}$ stabilizers are
referred as ``$e$'' type, while anyons corresponding to $\hat{B}$
stabilizers are ``$m$'' type.

\section{Interaction with the environment}
\label{sec:environment}

When the system is in contact with a bath the total Hamiltonian is
\begin{eqnarray}
  H = H_B + V_{BC},
  \label{eq:Hamilt}
\end{eqnarray}
where $H_B$ is the bath Hamiltonian and $ V_{BC}$ is the interaction
part. If the closed system is prepared in the product state $|SC
\rangle \otimes |B \rangle$, where the $|B\rangle$ is the bath ground
state, the closed system evolves in time as
\begin{eqnarray}
  |\psi(t) \rangle& =& U_I(t)\,|SC \rangle \otimes |B \rangle,
  \label{eq:productstate}
\end{eqnarray}
where $U_I(t)$ is the time evolution operator of the combined system
in the interaction picture. At the end of the QEC cycle the state of
the environment may have components beyond its ground state. As a
result, the entanglement between the qubits and the environment can
spill over to the next QEC cycle. While this effect deserves
investigation, here we will adopt a simplifying hypothesis and assume
that the environment remains in its ground state at the end of the QEC
cycle. This could be achieved by maintaining the environment cold
(i.e., by keeping it in contact with a much larger bath or reservoir). 
Hence, we define
\begin{eqnarray}
  U_{\rm eff}(\Delta) = \langle B|\, U_I (\Delta)\, |B \rangle,
  \label{eq:Ueff0}
\end{eqnarray}
as the effective evolution operator of the code system at the end of a
QEC cycle of duration $\Delta$. 

\begin{figure}[t]
  \includegraphics[width=0.8\columnwidth]{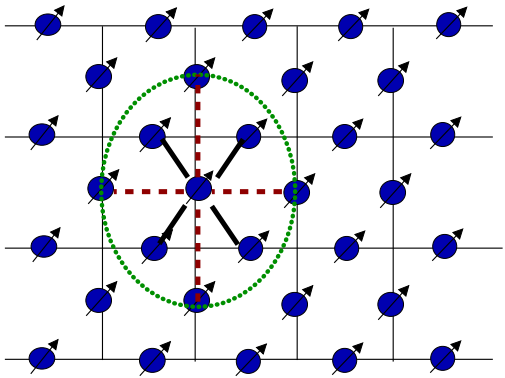}
  \caption{(Color online) The effective interaction induced by the
    bath, Eq. (\ref{eq:Ueff}), between a qubit and its nearest (solid
    black lines) and next-to-nearest neighbors (dashed red lines). The
    range of the interaction (dotted circle) is related to the QEC
    period $\Delta$. }
\label{fig:effectInteraction}
\end{figure}

The evolution under $U_{\rm eff}(\Delta)$ induces an effective
dynamics into the code system that includes dissipation and
dephasing. In general, the functional form of $U_{\rm eff}(\Delta)$ in
terms of the qubit operators $\{\sigma_i^\alpha\}$ can be rather
difficult to derive from first principles. For the particular case of
a gapless bosonic environment with a coupling given by
\begin{equation}
V_{BC} = \lambda \sum_{r_i} f(r_i)\, \sigma_i^x,
\end{equation}
a simple expression can be exactly derived. Here, $\lambda$ is the
strength of coupling to the bosonic field, $f(r_i)$ is the bosonic
field operator of the bath, and $\sigma_i^x$ is the Pauli matrix
acting on the $i$-th qubit. In this case, the induced evolution
operator dynamics was found to be \cite{novais2013,jouzdani2013}
\begin{equation}
U_{\rm eff}(\Delta) = e^{\beta H_{\rm eff}}=e^{-\beta\sum_{ij}J_{ij}
  \sigma^x_i\sigma^x_j}.
\label{eq:Ueff}
\end{equation}
The sum in the exponent is over the physical qubits of the surface
code (see Fig. \ref{fig:effectInteraction}). The parameter $\beta$ is
a function of $\lambda$ and other characteristics of the bosonic
environment. For the Ohmic case \cite{jouzdani2013},
\begin{equation}
\beta = \frac{1}{2\pi} \left( \frac{\lambda}{\omega_0} \right)^2,
\end{equation}
with $\omega_0$ denoting a characteristic frequency of the bosons and
\begin{equation}
\label{eq:Johmic}
J_{ij} = \frac{1}{2} \times \left\{ \begin{array}{cc} \mbox{arcosh}
  \left( \frac{v\Delta}{|{\bf r}_i-{\bf r}_j|} \right) +
  \frac{i\pi}{2}, & \frac{|{\bf r}_i - {\bf r}_j|} {v\Delta} < 1,
  \\ i\, \mbox{arcsin} \left( \frac{v\Delta} {|{\bf r}_i - {\bf r}_j|}
  \right), & \frac{|{\bf r}_i - {\bf r}_j|} {v\Delta} > 1. \end{array}
\right.
\end{equation}
Here $v$ is the bosonic mode velocity. The complex interaction
$J_{ij}$ is directly related to the correlation function of the bath
at two spacial points ${\bf r}_i$ and ${\bf r}_j$ and to
$\Delta$. Notice that Eq. (\ref{eq:Johmic}) was derived in
Ref. \cite{jouzdani2013} under the assumption that the bath returns to
its ground state at the end of the QEC cycle, as shown in
Eq. (\ref{eq:Ueff0}).
                                                                                   
The functional form in the effective action in Eq. (\ref{eq:Ueff}) can
also be used as a phenomenological error model of correlated errors
with a complex exchange coupling parameter $J_{ij}$. We note that
other forms of interaction between the surface code constituents and
the environment have been used in the literature. In particular, one
may start with an interaction such as $V_{BC} \sim \lambda
\sum_{r_{i}} f(r_{i})\, A_i$ where a bosonic field couples to the
stabilizers. In this case, the resulting effective dynamics may
enhance the surface code protection \cite{Pedrocchi13,Hamma08}.

\section{QEC with flawless recovery}
\label{QEC}

The system is maintained in its code space by means of QEC cycles. At
each QEC cycle the stabilizers are measured (syndrome detection) and a
suitable recovery operation is performed with the goal of returning
the system to its \emph{code state}, as defined in Sec. \ref{surf}.
Due to the interaction with the environment, prior to the syndrome
detection the code system is in a superposition state where all
syndromes are possible. After reading the stabilizers, the system is
detected (i.e., projected) onto a particular superposition state
(syndrome). Eventually, a nondestructive recovery returns the system
back to the code state, $|SC\rangle$, or erroneously to
$\hat{X}\,|SC\rangle$.

The syndromes in the QEC protocol are based on measurements of the
stabilizers. Let us define $\mathcal{P}_{n,f} = |n,f\rangle \langle
n,f|$ as the projection onto a subspace of $n$ excitations or
anyons. The $n$ excitations may be detected at different vertices or
plaquettes on the surface code. The index $f$ refers to the
configuration where anyons are detected on the surface code after
syndrome detection. Due to the topological nature of the code, any
state $|n,f\rangle$ is \emph{a superposition of a large number of
  states} involving the physical qubits of the system,
\begin{eqnarray}
  |n,f\rangle =  \sum_{s^*}   |s^*\rangle,
  \label{eq:syndrome}
\end{eqnarray}
where the sum is over configurations of the physical qubits
$|s^*\rangle = |s_1,\dots, s_N \rangle$ and the asterisk indicates
that the sum is taken over the configurations that are consistent with
the condition of ``$n$ anyons with the configuration $f$.''

After being initially set in the code state $|SC\rangle$, the system
interacts with the environment. After a time interval $\Delta$, it
evolves to the state $U_{\rm eff}(\Delta)\, |SC\rangle$. The effective
time evolution operator $U_{\rm eff}(\Delta)$ is in general nontrivial
and may not be unitary. At this point we assume that the QEC operation
detects the system (with some probability) to be in the state
$|n,f\rangle$. Then, a flawless recovery operation $\mathcal{R}$
returns this state to either $|SC\rangle$ or, erroneously, to
$\hat{X}|SC\rangle$, namely,
\begin{eqnarray}
  \mathcal{R} \, \mathcal{P}_{n,f}\, U_{\rm eff}(\Delta )\,|SC\rangle
  = \mathcal{A}\,|SC\rangle + \mathcal{B}\, \hat{X}\, |SC\rangle,
  \label{eq:recover}
\end{eqnarray}
where $\mathcal{A}$ and $\mathcal{B}$ are the amplitudes of the two
orthogonal states $|SC\rangle$ and $\hat{X}|SC\rangle$,
respectively. 

To be more explicit, let us exactly specify the projector
$\mathcal{P}_{n,f} = |n,f\rangle \langle n,f|$ for the case where
errors are of ``$m$'' type; ``$m$'' type errors occur as a result of
bit flipping qubits along a set of strings. We define this string
operation as
\begin{eqnarray}
  \hat{S}^x (\mathcal{L}) \, = \prod_{j\in \mathcal{L} } \sigma^x_j,
  \label{eq:String}
\end{eqnarray}
where $\mathcal{L}$ is a set of strings running on the surface code
lattice such that $|n,f\rangle = \hat {S}^x(\mathcal{L})
|SC\rangle$. There are many possible sets of $\mathcal{L}$ and
corresponding $ \hat {S}^x(\mathcal{L})$ that generate the same state
$|n,f\rangle$. Two such choices of string operators, $\hat
{S}^x(\mathcal{L}_1)$ and $\hat {S}^x(\mathcal{L}_2)$, can differ from
each other in two possible ways: Either
\begin{eqnarray}
  \hat {S}^x(\mathcal{L}_1) |SC\rangle \, &=& \left[ \prod_{s \in
      \mathcal{P}} \hat{A}_s \right] \, \hat {S}^x(\mathcal{L}_2)\,
  |SC\rangle \nonumber \\ &=& \hat {S}^x(\mathcal{L}_2)\, |SC\rangle,
\label{eq:StEquiv1}
\end{eqnarray}
or 
\begin{eqnarray}
  \hat {S}^x(\mathcal{L}_1) |SC\rangle \, &=& \left[ \prod_{s \in
      \mathcal{P}} \hat{A}_s \, \hat{X }\right] \hat {S}^x(\mathcal{L}_2)
  \,|SC\rangle \nonumber \\ &=& \hat{X}^\prime \hat
               {S}^x(\mathcal{L}_2)\, |SC\rangle.
  \label{eq:StEquiv2}
\end{eqnarray}
Here, $\mathcal{P}$ is a set of vertices on the surface code. In
Eqs. (\ref{eq:StEquiv1}) and (\ref{eq:StEquiv2}) we used the identity
$ \left[ \prod_{s \in \mathcal{P}} \hat{A}_s \right]\,|SC\rangle =
|SC\rangle$. Thus, the two states $S^x(\mathcal{L})\, |SC\rangle$ and
$\hat{X} S^x(\mathcal{L})\, |SC\rangle$ alone are enough to define
$\mathcal{P}_{n,f}$ as
\begin{eqnarray}
  \mathcal{P}_{n,f} &=& \hat {S}^x(\mathcal{L})\, |SC\rangle \langle
  SC| \,\hat {S}^x(\mathcal{L})\nonumber \\ &&+\, \hat{X} \hat
  {S}^x(\mathcal{L})\, |SC\rangle \langle SC | \, \hat
  {S}^x(\mathcal{L}) \hat{X}.
\label{eq:projc}
\end{eqnarray}
With this definition we find the state of the code at the end of the
first QEC cycle to be
\begin{eqnarray}
  | SC(\Delta) \rangle &=& \mathcal{R} \mathcal{P}_{n,f}\, U_{\rm
    eff}(\Delta )\, |SC\rangle \nonumber \\ & & + \, \mathcal{R} \hat
  {S}^x(\mathcal{L})\, |SC\rangle \,\, \langle SC| \hat
  {S}^x(\mathcal{L})\, U_{\rm eff}(\Delta )|SC\rangle \nonumber \\ &&
  + \, \mathcal{R}\, \hat{X}\, \hat {S}^x(\mathcal{L})\, |SC\rangle
  \,\, \langle SC | \hat {S}^x(\mathcal{L})\, \hat{X} U_{\rm
    eff}(\Delta )\,|SC\rangle \nonumber \\ &=& \mathcal{A}_{n,p}
  \,|SC\rangle + \mathcal{B}_{n,p} \hat{X}\, |SC\rangle,
  \label{eq:SCt}
\end{eqnarray}
where we assume a flawless recovery, $\mathcal{R}\, \hat
{S}^x(\mathcal{L}) = {\bf 1}$, and define the amplitudes
\begin{eqnarray}
\mathcal{A}_{n,f}= \langle SC| \hat {S}^x(\mathcal{L})\, U_{\rm
  eff}(\Delta )\,|SC\rangle
\label{eq:funcA}
\end{eqnarray}
and
\begin{eqnarray}
  \mathcal{B}_{n,f}&=&\langle SC | \hat {S}^x(\mathcal{L})\, \hat{X}\,
  U_{\rm eff}(\Delta )\,|SC\rangle \nonumber \\ &=& \langle SC | \hat
  {S}^x(\bar{ \mathcal{L}})\, U_{\rm eff}(\Delta )\,|SC\rangle.
\label{eq:funcB}
\end{eqnarray}

The fidelity is a suitable quantity to measure the success of the QEC
operation after error correction,
\begin{equation}
 \mathcal {F} = \frac{\langle SC |SC (\Delta) \rangle}
            {\sqrt{\langle SC (\Delta)|SC (\Delta) \rangle}}
\label{eq:fidel1}
\end{equation}
It is straightforward to show that the fidelity can be written in
terms of the amplitudes $\mathcal{A}_{n,f}$ and $\mathcal{B}_{n,f}$,
namely,
\begin{equation}
\mathcal {F} =\frac{1} {\sqrt { 1 + \left|
    \frac{\mathcal{B}_{n,f}}{\mathcal{A}_{n,f}} \right|^2}}.
\end{equation}

In order to find a suitable expression for the numerical evaluation of
the amplitudes $\mathcal{A}_{n,f}$ and $\mathcal{B}_{n,f}$, we write
$|SC\rangle$ as
\begin{equation}
|SC\rangle = \frac{1}{\sqrt{2^{N_\diamondsuit}}} \prod_{s}{(1+ \hat{A}_s)}\,
|F_z\rangle,
\label{eq:SC1}
\end{equation}
where $|F_z\rangle$ is the ferromagnet state of the qubits in the
$z$-direction: $|F_z\rangle = |\uparrow\rangle_1 \dots
|\uparrow\rangle_N$, with $N_\diamondsuit$ being number of star
operators. Noting that
\begin{equation}
|F_z\rangle = \prod_{i=1}^{N} \frac{|+\rangle_i\, + \,
  |-\rangle_i}{\sqrt{2}},
\end{equation}
where $\sigma_i^x\,|\pm\rangle_i= \pm1\,|\pm\rangle_i$ and
$|\pm\rangle_i$ stands for eigenvectors of Pauli matrix $\sigma^x_i$
acting on $i$-th physical qubit, we obtain
\begin{equation}
|SC\rangle = \frac{1}{\sqrt{2^{N_\diamondsuit-1}}}
\sum_{s^*}|s^*\rangle.
\label{eq:SC2}
\end{equation}
The sum in Eq. (\ref{eq:SC2}) runs over restricted states $s^*$ (a
product state of $|\pm\rangle_i$ of physical qubits) that preserve the
conditions $A_s=1$ (i.e., $\hat{A}_s|s^*\rangle=+|s^*\rangle$) for all
vertices $s$ of the lattice. The state $|SC\rangle$ also satisfies the
conditions $\hat{Z}|SC\rangle=+|SC\rangle$ and $B_p=1$
($\hat{B}_p|SC\rangle=+|SC\rangle$) for all plaquettes of the
lattice. Hereafter we will make use of the relations,
\begin{equation}
 \hat {S}^x(\mathcal{L}) |s^*\rangle = S_{s^\star}(\mathcal{L})\,
 |s^*\rangle,
  \label{eq:AS}
\end{equation}
and 
\begin{equation}
 \hat {S}^x(\mathcal{L}) \hat{X} |s^*\rangle = S_{s^\star}(\bar
      {\mathcal{L}})\, |s^*\rangle,
  \label{eq:BS}
\end{equation}
with $S_{s^\star}= \pm1$ being the product of the $\sigma^x_i$
operators along either the path $\mathcal{L}$ or $\bar {\mathcal{L}}$.
By inserting Eqs. (\ref{eq:SC2})-(\ref{eq:BS})
into Eqs. (\ref{eq:funcA}) and (\ref{eq:funcB}), we arrive at
\begin{eqnarray}
 \mathcal{B} = \frac{\mathcal{B}_{n,f}}{\mathcal{A}_{0,0}} &=&
 \frac{\sum_{s^*}S_{s^\star}(\bar{\mathcal{L}}) \, U_{\rm
     eff}(s^\star)}{\sum_{s^*} \, U_{\rm eff}(s^\star) },
\label{eq:expX}
\end{eqnarray}
and 
\begin{eqnarray}
 \mathcal{A}= \frac{\mathcal{A}_{n,f}}{\mathcal{A}_{0,0}} &=&
 \frac{\sum_{s^*}S_{s^\star}(\mathcal{L}) \, U_{\rm
     eff}(s^\star)}{\sum_{s^*} \, U_{\rm eff}(s^\star) }.
\label{eq:partition}
\end{eqnarray}
Here $U_{\rm eff}(s^\star)$ is the matrix element $\langle s^\star |
U_{\rm eff}(\Delta ) | s^\star \rangle$ that can be considered as a
statistical weight in the sums shown above.

Equations (\ref{eq:expX}) and (\ref{eq:partition}) show that the
calculation of $\mathcal{A}_{n,f}$ and $\mathcal{B}_{n,f}$ maps onto a
\emph{statistical mechanics} problem where these amplitudes are equal
to the expectation values $\langle S(\mathcal{L}) \rangle$ and $
\langle S(\bar{\mathcal{L}}) \rangle$, respectively. The averaging
$\langle \dots \rangle$ is performed with respect to a complex-time
effective action $H_{\rm eff}$ that gives rise to the statistical
weight $U_{\rm eff}(s^\ast)$ introduced above. In the following, we
study the fidelity of the code based on the behavior of the amplitudes
$\mathcal{B}$ and $\mathcal{A}$ for an effective action comprising
qubit-qubit interactions of the form introduced in
Eq. (\ref{eq:Ueff}). We limit our study to real values of $\beta$ and
$J_{ij}$ while probing different ranges of interactions, namely
nearest neighbors and next-to-nearest neighbors. We remark that the
range of the effective qubit-qubit interaction can be sharply
controlled by the duration of the QEC cycle: Longer cycles lead to
longer ranges while shorter cycles decrease the range, even down to
nearest neighbors. 

\section{Numerical Method}
\label{sec:method}

We numerically evaluate the amplitudes $\mathcal{A}$ and
$\mathcal{B}$, as defined in Eqs.  (\ref{eq:expX}) and
(\ref{eq:partition}), using a classical Monte Carlo method and
assuming an effective evolution operator as in Eq. (\ref{eq:Ueff}),
with an effective action of the form,
\begin{equation}
H_{\rm eff}= - \sum_{(ij)} \,J(|r_i-r_j|)\, {\sigma^x_i\sigma^x_j}
\end{equation}
for nearest-neighbor and beyond nearest-neighbor interactions.

Using the standard classical Monte Carlo method \cite{Kardar07}, we
replace the summation over the large set of configurations $\{s^*\}$
in Eqs. (\ref{eq:expX}) and (\ref{eq:partition}), which is of order
$[O(2^{\frac{N}{2}})]$, by a sum over a smaller set of representative
sample configurations $\{ \tau \}$ for a given value of $\beta J$. All
the sampled configurations have the same statistical weight $e^{-\beta
  E_{\tau}}$. If there are $M$ representative configurations for a
given $\beta J$, we then have
\begin{equation}
  \sum_{s^*} \{...\} \,e^{-\beta E_{s^*}} \rightarrow
  \frac{1}{M}\sum_{\tau} \{...\},
  \label{eq:MC1}
\end{equation}
where we target the average value of the quantity $\{...\}$. We use a
Metropolis algorithm to collect these relevant configurations assuming
that $\beta J$ is real. However, since we must take into account the
constraint $A_s=1$ for any vertex $s$, the standard Metropolis
algorithm needs to be suitably modified.

In general, a state $|s^* \rangle$ that satisfies the constraint
``$A_s=1$ for all $s$'' in Eq. (\ref{eq:SC2}) has the form,
\begin{equation}
|s^* \rangle = \begin{cases} \prod_{p \in \mathcal{P}} \,\hat{B}_p \,
  |F_x\rangle,& \text{(I)}\\ \text{or}&\\ \prod_{p \in \mathcal{P}}
  \,\hat{B}_p \, \hat{Z}\, |F_x\rangle. &\text{(II)}
  \end{cases}
  \label{eq:MC2}
\end{equation}
Here, $\mathcal{P}$ is an arbitrary set of plaquettes. States in the
first class, $(I)$, are eigenstates of $\hat{X}$ with eigenvalue $+1$,
while in the second class, $(II)$, the states are eigenstates of
$\hat{X}$ with eigenvalue $-1$. Equation (\ref{eq:MC2}) provides a
natural codification of the restricted states $|s^* \rangle$: One can
start with a \emph{vacuum state} $|F_x\rangle$, then flip a number of
qubits by $\prod_{p \in \mathcal{P}} \,\hat{B}_p$ that correspond to
$\mathcal{P}$, and arrive at a restricted state $|s^* \rangle$.
However, this is not the route we pursue. Below we present an
equivalent but alternative definition for the stabilizers of the
surface code and consequently of the states in
Eq. (\ref{eq:MC2}). They provide a more efficient implementation of
the Metropolis algorithm. The alternative definition for stabilizers is not
limited to the surface code and can be extended to higher-dimensional
codes.

\begin{figure}[t]
\includegraphics[width=0.8\columnwidth]{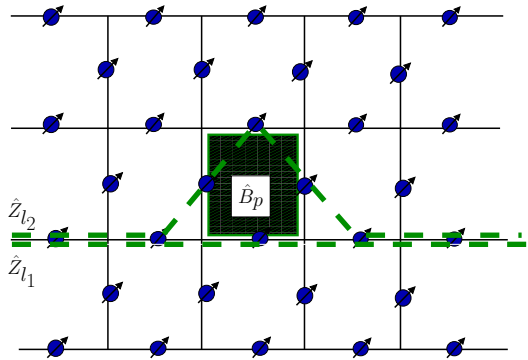}
\caption{(Color online) Applying two logical $\hat{Z}$ operators along
  the paths $l_1$ and $l_2$ is equivalent to a $\hat{B}_p$ operator
  shown with the hatched rectangle.}
\label{fig:StabilizersDefinition}
\end{figure}

Since $\sigma_i^\alpha \sigma_i^\alpha =1$ for $\alpha=x,y,z$, one can
write the stabilizers of Eqs. (\ref{eq:StabilizersA}) and
(\ref{eq:StabilizersB}) as
\begin{eqnarray}
\hat{A}_s = \prod_{i\in s} \sigma_i^x = \hat{X}_{l_1} \hat{X}_{l_2}
\dots \hat{X}_{l_{2m}}
  \label{eq:MC31}
\end{eqnarray}
and
\begin{eqnarray}
\hat{B}_p = \prod_{i\in p} \sigma_i^z = \hat{Z}_{l_1} \hat{Z}_{l_2}
\dots \hat{Z}_{l_{2m}},
 \label{eq:MC32}
\end{eqnarray}
where a path $l_{i}$ goes from one boundary to the opposite boundary
of the system. The set $\{l_i\}$ is chosen such that the product of
the logical operators $\hat{X}_{l_1} \hat{X}_{l_2} \dots
\hat{X}_{l_{2m}}$ (or $\hat{Z}_{l_1} \hat{Z}_{l_2} \dots
\hat{Z}_{l_{2m}}$) is equal to $\prod_{i\in s} \sigma_i^x$ (or
$\prod_{i\in p} \sigma_i^z$) and thus the stabilizer $\hat{A}_s$ (or
$\hat{B}_p$). The number of paths, $2m$, is not unique. An even number
guarantees that the commutation relation $[\hat{A}_s,\hat{B}_p]=0$ is
satisfied. The product of $2m$ logical operators always forms closed
loops. For example, in Fig. \ref{fig:StabilizersDefinition} the
operation of the stabilizer operator $\hat{B}_p$ on the qubits of
plaquette $p$ is generated by applying two logical $\hat{Z}$ operators
along the paths $l_1$ and $l_2$, as depicted in the figure.

Thus, the states $|s^*\rangle$ defined in Eq. (\ref{eq:MC2}) can be
stated in terms of logical operations as
\begin{equation}
|s^* \rangle = \begin{cases} \hat{Z}_{l_1} \hat{Z}_{l_2} \dots
  \hat{Z}_{l_{2m}} \, |F_x\rangle = {\bf Z}_{{\it l}}^2 \,
  |F_x\rangle, & \text{(I)}\\ \text{or}&\\ \hat{Z}_{l_1} \hat{Z}_{l_2}
  \dots \hat{Z}_{l_{2m}} \hat{Z}|F_x\rangle = {\bf Z}_{{\it l}}
  |F_x\rangle,& \text{(II)}
  \end{cases}
  \label{eq:MC4}
\end{equation}
where we abbreviate the product $\hat{Z}_{l_1} \hat{Z}_{l_2} \dots
\hat{Z}_{l_{2m}}$ as ${\bf Z}_{{\it l}}^2 $ and $\hat{Z}_{l_1}
\hat{Z}_{l_2} \dots \hat{Z}_{l_{2m}} \hat{Z}$ as ${\bf Z}_{{\it l}}$.
By introducing these definitions into Eqs. (\ref{eq:expX}) and
(\ref{eq:partition}) and using the fact that ${\bf Z}_{{\it l}}^2 \,
\hat{X}= \hat{X}\, {\bf Z}_{{\it l}}^2$ and ${\bf Z}_{{\it l}} \,
\hat{X}= -\, \hat{X}\, {\bf Z}_{{\it l}}$, we arrive at
\begin{eqnarray}
  \mathcal{B} &=& \sum_{{\it l}} \{ \langle F_x|\, {\bf Z}_{{\it
      l}}^2\, S^x(\mathcal{L})\,U_{\rm eff}\,{\bf Z}_{{\it l}}^2
  |F_x\rangle \nonumber \\ & &-\ \langle F_x|\, {\bf Z}_{{\it l}}\,
  S^x(\mathcal{L}) \,U_{\rm eff}\,{\bf Z}_{{\it l}} |F_x\rangle\}
\label{eq:MC51}
\end{eqnarray}
and 
\begin{eqnarray}
\mathcal{A} &=& \sum_{{\it l}} \{\langle F_x|\, {\bf Z}_{{\it l}}^2\,
S^x(\mathcal{L})\,U_{\rm eff}\,{\bf Z}_{{\it l}}^2 |F_x\rangle
\nonumber \\ & & +\ \langle F_x|\, {\bf Z}_{{\it
    l}}\,S^x(\mathcal{L})\,U_{\rm eff}\,{\bf Z}_{{\it l}}
|F_x\rangle\},
\label{eq:MC52}
\end{eqnarray}
up to a common normalization factor.

To understand the essential difference between $\mathcal{A}$ and
$\mathcal{B}$, let us assume a phase of the system in which the states
belonging to the two classes of Eq. (\ref{eq:MC4}) contribute with the
same statistical weight $U_{\rm eff}$ (the topological state).  In
this phase any fluctuation around the equilibrium configuration
states, $ \{|s^\star \rangle\}$ (which is of the order of
$\mathcal{L}$ and less than the distance of the code $\frac{L}{2}$),
will be canceled out in the sum in the expression for $\mathcal{B}$
via the minus sign of the second term in Eq. (\ref{eq:MC51}). Hence,
the ratio $\left| \frac{\mathcal{B}}{\mathcal{A}} \right| \rightarrow
0$ in the thermodynamic limit and the fidelity $\mathcal{F}\rightarrow 1$,
as expected. However, in the phase where the statistical weight of the
states in class $I$ differs from states in class $II$ in
Eq. (\ref{eq:MC4}), i.e., in the ordered phase, there is a sizable
change in $\left| \frac{\mathcal{B}}{\mathcal{A}} \right|$. In our
model the ordered phase corresponds to the state $|F_x\rangle$. In
this limit, $\left| \frac{\mathcal{B}}{\mathcal{A}} \right|\rightarrow
1$ and a sharp phase transition takes place between these two
limits. Thus, for a correct decoding and sufficiently large system,
one should expect to see $|\mathcal{B}| < |\mathcal{A}|$ in the
disordered phase (topological phase) in a universal way, independently
of the error $S_{s^\star}(\mathcal{L})$, as long as $\mathcal{L} <
\frac{L}{2}$.

Equation (\ref{eq:MC4}) provides a novel way for the codification of
the restricted state $|s^* \rangle$: One begins with a \emph{vacuum
  state} $|F_x\rangle$, then flips a number of qubits along a certain
path $\{l \equiv l_1\dots\}$, and arrives at a restricted state ${\bf
  Z}_{l}^2 \, |F_x\rangle$ or ${\bf Z}_{{l}} \, |F_x\rangle$. The sums
in Eqs. (\ref{eq:MC51}) and (\ref{eq:MC52}) run over all possible
paths $\{l\}$. The statistical weight $U_{\text{eff}}$ corresponds to
the probability of flipping the qubits along the path ${\it l}$. In
this regard, the Metropolis algorithm finds the most relevant paths.
The scheme to update the configurations is then similar to the
techniques used in world-line-based quantum Monte Carlo
\cite{Assad2006}, since each two ${\bf Z}_{l}$ and ${\bf
  Z}_{l^\prime}$ differ by a certain number of $\hat{B}_p$ operations.

\section{Numerical Results}
\label{sec:results}

Considering a real homogeneous interaction coupling $J_{ij}=
J(|r_i-r_j|)$, we use the method described in Sec. \ref{sec:method} to
numerically evaluate $\mathcal{A}$, $\mathcal{B}$, and $\left|
\frac{\mathcal{B}}{\mathcal{A}} \right|$ to determine the fidelity
$\mathcal{F}$.

\begin{figure}[h]
  \includegraphics[width=0.8\columnwidth]{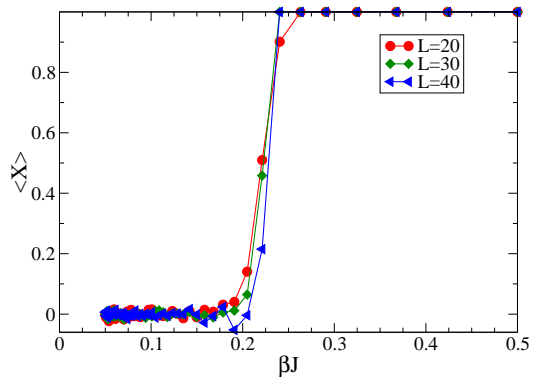}
  \caption{(Color online) The numerical evaluation of $\langle X
    \rangle = \left| \frac{\mathcal{B}}{\mathcal{A}} \right|$ for
    non-error syndromes based on the Monte Carlo calculation for
    different system sizes. $L=20$ is a surface code system with $761$
    qubits (circle), $L=30$ has $1741$ qubits (diamond), and $L=40$
    has $3121$ qubits. The solid lines are guides to the eye. On the
    horizontal axis, $\beta$ is proportional to the coupling to the
    environment and $J$ is the exchange coupling of the effective
    interaction between nearest-neighbor qubits. In this simulation
    80,000 iterations are used for each $\beta$ step.}
\label{fig:result1}
\end{figure}

\subsection{Results for the non-error sector $\mathcal{P}_{0,0}$}

For the no-charge sector $\mathcal{P}_{0,0}$ we have
$S^x(\mathcal{L})={\bf 1}$. Our numerical results show that $\left|
\frac{\mathcal{B}}{\mathcal{A}}\right|$ follows the local order
parameter $\langle \sigma_i^x \rangle$, where $i$ is an arbitrary
qubit in the bulk of the surface code system. The behavior of
$\left|\frac{\mathcal{B}}{\mathcal{A}}\right|$ as a function of $\beta
J$ for nearest-neighbor interaction and different system sizes is
shown in Fig. \ref{fig:result1}.

\begin{figure}[b]
\includegraphics[width=0.8\columnwidth]{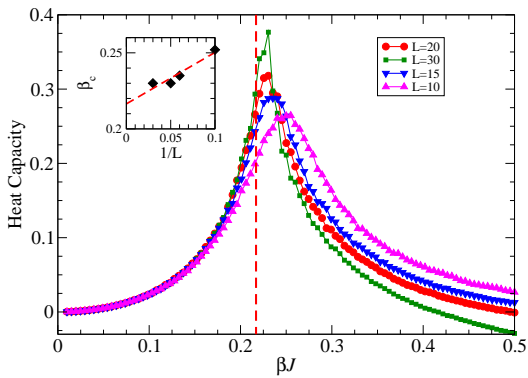}
\caption{(Color online) Finite-size scaling analysis of the heat
  capacity per qubit that yields $\beta_c = 0.217$ for $L \rightarrow
  \infty$. The solid lines are guides to the eye.}
\label{fig:result2}
\end{figure}

The surface code geometry can be decomposed into two sublattices. Here
we considered square sublattices of sizes $L\times L$ and $(L-1)\times
(L-1)$. By increasing the system size the transition from the
topological state, where $\beta J<\beta_cJ$, to a trivial state where
the degeneracy between $|s^*\rangle$ states is lifted, becomes
sharper. This confirms the first-order phase transition nature of the
effect (i.e., the existence of an error threshold in the fidelity). A
finite-size scaling of the heat capacity is shown in
Fig. \ref{fig:result2}. Taking $\beta$ as a fictitious inverse
temperature, we used $\frac{\beta^2(\langle E^2 \rangle - \langle E
  \rangle^2)}{V}$ as the definition of heat capacity. Here $V$ is the
total number of qubits. By setting $J=1$ and fitting the data to the
asymptotic functional form $\beta_c(L) = \beta_c(\infty) - y L^x$, we
find the critical exponent $x=-1/\nu = -1$, in agreement with the
expected Ising model ($\nu=1$). Scaling the data according to this
functional form also gives $\beta_c(\infty) = 0.217$. This value
agrees closely with the analytical result obtained in
Ref. \cite{jouzdani14}.

\begin{figure}[t]
\includegraphics[width=0.8\columnwidth]{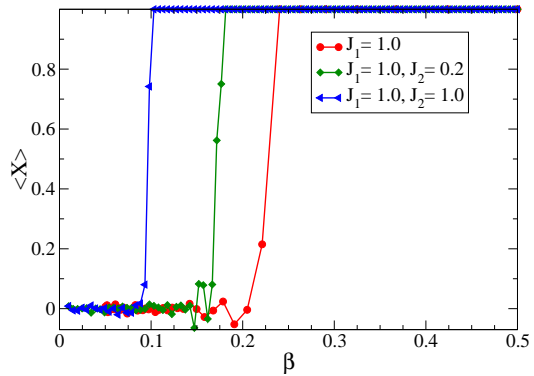}
\caption{(Color online) The ratio $\left| \frac{\mathcal{B}}
  {\mathcal{A}} \right|$ for different interaction ranges as a
  function of $\beta$ on a lattice with $L=40$ and 80,000 Monte Carlo
  steps for each data point. The data points correspond to $J_1=1$
  (circles), as in Fig. \ref{fig:result1}; $J_1=1$, $J_2=0.2$, and
  $J_m=0$ for $m>2$ (diamonds); and $J_1=1$, $J_2=1$, and $J_m=0$ for
  $m>2$ (triangles) in Eq. (\ref{eq:effectInteractionrange}). The
  solid lines are guides to the eye.}
\label{fig:result3}
\end{figure}

In this one-cycle study, the time period of the cycle, $\Delta$,
affects directly the range of interaction in $H_{\rm eff}$ [see, for
  example, Eq. (\ref{eq:Johmic})], while the strength is proportional
to $\beta$. Keeping the environment parameters fixed, the longer the
error cycle, the longer the range of the interactions, as qubit
correlations are intermediated by the propagation of environmental
modes. To extend the analysis to qubit-qubit interactions beyond
nearest neighbors, we write
\begin{eqnarray}
  H_{\rm eff} &=& - \sum_m\sum_{m\text{-th n.n }(ij)} J_m\,
  \sigma^x_i\, \sigma^x_j,
\label{eq:effectInteractionrange}
\end{eqnarray}
where $J_m$ is the exchange coupling between $m$-th nearest neighbors
(see Fig. \ref{fig:effectInteraction}). In Fig. \ref{fig:result3} the
behavior of $\left| \frac{\mathcal{B}}{\mathcal{A}} \right|$ as a
function of $\beta$ is shown for some fixed values of $J_m$. By
increasing $\Delta$ the interaction range in $H_{\rm eff}$ varies and
therefore one needs to take into account higher orders of $m$ in
Eq. (\ref{eq:effectInteractionrange}). By including higher orders of
$m$, the threshold value in the coupling to the environment for which
the code protection is lost also changes. We see that a longer QEC
cycle brings a larger range of correlated errors into account and
consequently decreases the threshold value $\beta_c$. This indicates
that for increasing values of $\Delta$, a smaller coupling to the bath
is sufficient to destroy the topological state of the surface code. We
should emphasize that the effect is robust against increases of system
size and the value of $\beta_c$ is also well defined in the
thermodynamic limit in this case. In general, the dependence of $J_m$
on $\Delta$ is determined by the characteristics (correlation
functions) of the environment; for bosonic environments this
dependence was derived for some representative cases in
Ref. \cite{jouzdani2013}. We have numerically calculated the ratio
$\left| \frac{\mathcal{B}}{\mathcal{A}} \right|$ for interaction
ranges up to the fourth nearest neighbor. The results (not shown)
confirm a trend of decreasing thresholds when the interaction range is
increased.

\subsection{Results for one-error sector $\mathcal{P}_{1,f}$ and the two-error sector $\mathcal{P}_{2,f}$}

\begin{figure}[t]
\includegraphics[width=0.85\columnwidth]{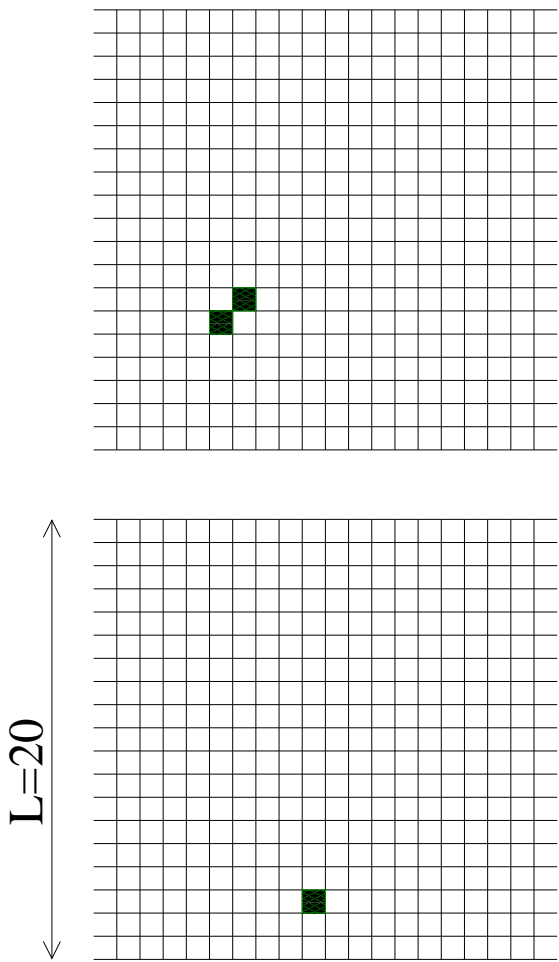}
\caption{(Color online) The location of the single error $\{B_{p_0}\}$
  (bottom) and two errors $\{B_{p_1},B_{p_2}\}$ (top) assumed in the
  numerical calculations.}
\label{fig:errors}
\end{figure}

To investigate the intrinsic nature of the transition mentioned above
we have also numerically evaluated $ \langle S^x_{1,2}\rangle = \left|
\frac{\mathcal{B}}{\mathcal{A}} \right|$ for charge sectors
$\mathcal{P}_{1,f}$ (where a plaquette $B_{p_0}$ is measured to be
$-1$) and $\mathcal{P}_{2, f}$ (where two plaquettes $B_{p_1}$ and
$B_{p_2}$ are measured $-1$). The locations of the errors
$\{B_{p_0}\}$ and $\{B_{p_1},B_{p_2}\}$ are arbitrarily chosen as
shown in Fig. \ref{fig:errors}.

\begin{figure}[b]
\includegraphics[width=0.80\columnwidth]{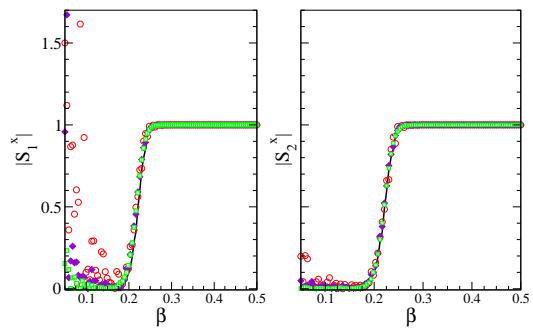}
\caption{(Color online) The ratio $|\frac{\mathcal{B}}{\mathcal{A}}|$
  for a lattice of $L=20$. The left box shows the convergence of $
  \langle S^x_{1}\rangle= |\frac{\mathcal{B}}{\mathcal{A}}|$ for one
  detected error located as shown in Fig. \ref{fig:errors}. The right
  box shows the convergence of $ \langle S^x_{2}\rangle =
  |\frac{\mathcal{B}}{\mathcal{A}}|$ for two detected errors located
  as shown in Fig. \ref{fig:errors}. In both boxes the number of
  iterations used for each $\beta$ data point is: $90,000$ (circle),
  $180,000$ (square), and $900,000$ (diamonds). The solid line is
  obtained from the corresponding no-error sector
  (Fig. \ref{fig:result3}).}
\label{fig:errors1}
\end{figure}

In the presence of detected errors, the numerical calculations a require
higher number of iterations. Figure \ref{fig:errors1} shows the
gradual convergence of the results for the one- and two-error sectors
to the results achieved for the no-error sector. In these calculations
only the nearest-neighbor case ($J_1=1$) has been considered. As can
be seen in Fig. \ref{fig:errors1}, for small values of $\beta$, 
complete convergence is not achieved when the number of iterations
is just $O(10^4)$ per data point and a much larger data 
set is required. However, the data shows a clear
tendency of convergence toward the same curve obtained for the
no-error sector when the number of iterations is increased. The
results for the case with a larger range of correlated errors [$J_m
  \ne 0$ for $m>1$ in Eq. (\ref{eq:effectInteractionrange})] converge
toward their counterpart of no-error syndrome as well (not shown). In
fact, we observe a faster convergence when the range of correlations
is larger. Results for other nonzero error configurations different
than the configurations considered here were found to be consistence
with the results shown in Fig. \ref{fig:errors1}. However, a larger
distance between errors requires a significantly higher number of
computational iterations to achieve convergence.

\section{Summary}
\label{sec:summary}

A non-Markovian environment in contact with the surface code induces
an effective dynamics (action) on the code system. Environmental
degrees of freedom can intermediate interactions between physical
qubits making up the system. As a result, when errors occur during the
code evolution, they will be correlated. We considered the effect of
such correlated errors on the fidelity of the code state after one
error correction cycle. We studied the code state resulting from an
effective action derived for a gapless bosonic environment but expect
the same model to describe phenomenologically other types of
environments. The calculation of the expectation values that enter in
the fidelity can be cast in the form of expectation values of a
statistical mechanics spin model with two separate phases. The
disordered and ordered phases of the statistical version correspond to
the topological and nontopological states of the surface code system
in contact with the environment. We evaluated an upper bound for the
threshold of the coupling to the bath beyond which no quantum error
correction is possible (i.e., fidelity is fully lost).

We showed numerically that the transition between the two phases can
be evaluated by a Monte Carlo method. We used a definition for the
stabilizers of the code based on the logical errors. The logical error
in this definition plays a role equivalent to a world-line in the
world-line-based quantum Monte Carlo. The separation of the two phases
of the surface code lies behind the statistical physics of these
world-lines, as presented in Eqs. (\ref{eq:MC51}) and
(\ref{eq:MC52}). The alternative definition for stabilizers given in
this paper can be extended to higher-dimensional topological codes
where the stabilizers are defined on hypercubes and logical errors
correspond to closed hypersurfaces \cite{Castelnovo08}. In higher-
dimensional codes the stabilizers can be defined in terms of the
logical errors similar to Eqs. (\ref{eq:MC31}) and
(\ref{eq:MC32}). Therefore, a similar approach should be applicable to
those codes.

In the numerical evaluation we considered qubit-qubit interactions
with different interaction ranges. We considered a QEC cycle with both
zero and nonzero error syndromes. For the nearest-neighbor range the
results perfectly agrees with the analytical calculations in
Ref. \cite{jouzdani14}. Finite-size scaling shows the value for the
threshold $\beta_c$ to be close to half of an Ising model with
nearest-neighbor interaction. For a longer range of interactions the
threshold $\beta_c$ decreases. The type of error syndrome does not
affect the value $\beta_c$. However, higher numerical iterations were
required to achieve convergence beyond nonerror syndromes.

Our results are based on the assumption that the interaction between
the physical qubits and the environment has the form $V_{BC} = \lambda
\sum_{r_i} f(r_i)\, \sigma_i^x$. Different functional forms for this
interaction may result in a different effective evolution operator
$U_{\rm eff}$ than the one studied here and may set different
threshold values for the fidelity. Another question that should be
addressed is the behavior of the fidelity over multiple QEC
cycles. Both issues are open to future investigations.

\begin{acknowledgments}
We thank Robert Raussendorf, Igor Tupitsyn, and Eduardo Novais for
insightful conversations. This work was supported in part by the
office of Naval Research and the National Science Foundation under
Grant No. CCF-1117241.
\end{acknowledgments}

\end{document}